\documentclass[dvips]{article}
\usepackage{icrctc07}

\title{The inefficiency of the first-order Fermi process in UHECR production 
at relativistic shocks}
\shorttitle{Cosmic-ray acceleration at relatvistic shocks}
\authors{J. Niemiec$^{1,2}$, M. Ostrowski$^{3}$, M. Pohl$^{1}$.}
\shortauthors{Niemiec and et al}
\afiliations{$^1$Department of Physics and Astronomy, Iowa State University, Ames, 
IA 50011, USA\\ $^2$Instytut Fizyki J\c{a}drowej PAN, ul. Radzikowskiego 152,
 31-342 Krak\'{o}w, Poland\\ $^3$Obserwatorium Astronomiczne, Uniwersytet Jagiello\'{n}ski,
ul. Orla 171, 30-244 Krak\'{o}w, Poland}
\email{niemiec@iastate.edu}

\abstract{The question of the origin of ultra-high-energy cosmic rays at 
relativistic shock waves is discussed in the light of results of recent Monte 
Carlo studies of the first-order Fermi particle acceleration 
\cite{nie06a,nie06b}. The models of the 
turbulent magnetic field near the shock considered in these simulations include 
realistic features of the perturbed magnetic field structures at the shock, 
which allow us to study all the field and particle motion characteristics that 
are important for cosmic-ray acceleration. Our results show that turbulent 
conditions near the shock, that are consistent with the shock jump conditions, 
lead to substantial modifications of the acceleration process with respect to 
the simplified models, that produce wide-range power-law energy distributions, 
often with the "universal" spectral index. Relativistic shocks are essentially 
always superluminal, and thus they preferentially generate steep particle 
spectra with cutoffs well below the maximum scattering energy, often not 
exceeding the energy of the compressed background plasma ions. Thus, cosmic-ray 
acceleration to very high energies at relativistic shock waves is inefficient, 
and such shocks are not expected to be the sources of ultra-high-energy 
particles.}

\begin{document}
\maketitle
%Begin the section.

\subsection{Introduction}
Relativistic shocks are widely considered to generate energetic particle
populations (cosmic rays) responsible for the high-energy emission of astrophysical
sources such as hot spots in radio galaxies, quasar jets and gamma-ray burst
afterglows. The basic acceleration mechanism discussed in this context is the
first-order Fermi process. It is believed that the Fermi process is intrinsically
efficient and thus also capable of the production of ultra-high-energy particles. 
In this work we confront this opinion and show that the generation of UHECRs
at relativistic shocks must invoke processes other that the first-order Fermi
mechanism.%  
\begin{figure}[t]
  \includegraphics[height=.27\textheight]{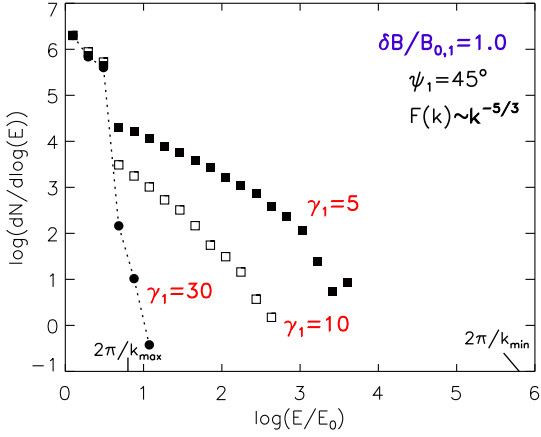}
  \caption{Accelerated particle spectra in the shock rest frame
  at oblique ($\psi_1=45^o$) superluminal 
shock waves for different shock Lorentz factors $\gamma_1$. The Kolmogorov wave 
power spectrum is assumed for the turbulent magnetic field, and the upstream
perturbation amplitude $\delta B / B_{0,1}=1.0$. Particles in the range 
($2\pi/k_{max}$, $2\pi/k_{min}$) can satisfy the resonance condition 
$k_{res}\simeq 2\pi /r_g(E)$ for some of the waves in the turbulence spectrum.}
\end{figure}
\subsection{Numerical models and results}
Modeling of first-order Fermi acceleration at relativistic shocks is a difficult 
task because cosmic-ray distributions are highly anisotropic at the shock and the
resulting particle spectra depend strongly on the essentially unknown local 
conditions at the shocks. In the series of our recent studies  of the Fermi process
\cite{nie04,nie06a,nie06b} we have to considered the most realistic models 
possible 
for the perturbed magnetic field structures at the shock, which allow us to study 
all the field characteristics important for particle acceleration. 
The upstream magnetic field is assumed to consist of the 
uniform component {\bf\em B}$_{0,1}$,
inclined at an angle $\psi_1$ to the shock normal%
\footnote{Indices "1" and "2" refer to quantities in the upstream and downstream 
plasma rest frame, respectively.}, 
and static finite-amplitude 
perturbations imposed upon it. The irregular component has 
either a flat $(F(k)\sim k^{-1})$ or a Kolmogorov $(F(k)\sim k^{-5/3})$ wave 
power spectrum defined in a wide wavevector range with $k_{max}/k_{min}=10^5$,
which allows us to investigate the role of the long-wave turbulence in the
acceleration process.  
The downstream field structure is obtained as the compressed upstream field,
so that the magnetic field lines are continuous across the shock. This allows 
one to study upstream-downstream correlations in particle motion introduced
by the field structure for different levels of turbulence, and to investigate 
the influence of this factor on particle spectra. 
We study the first-order Fermi process in {\em test-particle} approach with the method 
of Monte Carlo simulations, which calculates the particle spectra by following 
exact particle trajectories in the perturbed magnetic field near the shock.
A shock has a planar geometry and propagates with Lorentz factor $\gamma_1$ 
with respect to the upstream plasma.

Because nearly all magnetic field configurations in relativistic shocks 
lead to a perpendicular (superluminal) shock structure, the characteristic 
features of particle acceleration processes at high-$\gamma$ shocks are best 
illustrated using the oblique shock example of Fig. 1.   
All injected 
particles are initially accelerated in a phase of ``superadiabatic'' compression 
at the shock \cite{beg90}. Only a much smaller fraction of these particles is further 
accelerated in the first-order Fermi process, forming an energetic tail in the 
spectrum for highly perturbed magnetic fields. The shape of the spectral tail
and its extension to high particle energies strongly depend on the magnetic 
field turbulence spectrum. The tails for the Kolmogorov turbulence (Fig. 1), with 
most power in long-wave perturbations, are much flatter than for the flat wave 
power spectrum. However, in either case, the spectra steepen and/or the energy
cut-offs occur in the resonance 
energy range, and the cut-off energy decreases with growing shock 
Lorentz factor. 
\begin{figure}
  \includegraphics[height=.27\textheight]{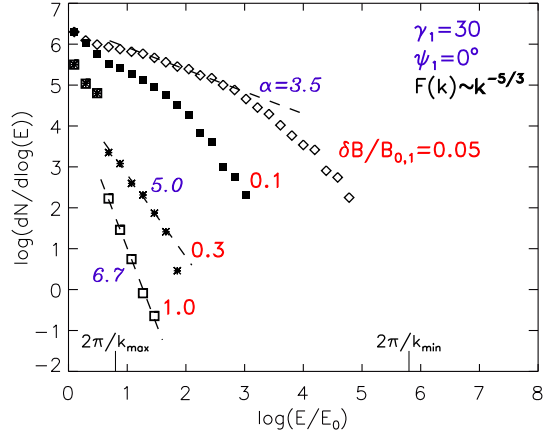}  
  \caption{Particle spectra at parallel shock waves with $\gamma_1$=30 for 
  different amplitudes of the magnetic field perturbations $\delta B / B_{0,1}$.
  Linear fits to the spectra are presented and values
  of the (phase-space) spectral indices $\alpha$ are given in italic 
(the energy spectral index $\sigma=\alpha -2$).
Some spectra are vertically shifted for clarity.}
\end{figure}
These spectral features result from the character of particle transport 
in the magnetic field downstream of the shock, where field compression 
produces effectively 2D turbulence, in which particle diffusion along 
the shock normal is strongly suppressed. In effect, advection of particles 
with the downstream flow leads to high particle escape rates, resulting in steep 
particle spectra.
The existence of the Kolmogorov turbulence at the shock allows for the formation 
of more extended and flatter spectral components, due to the effects of 
high-amplitude long-wave magnetic field perturbations which can form locally 
subluminal field configurations at the shock, thus enabling more efficient 
particle-shock interactions. However, the importance of these effects diminishes
for larger shock Lorentz factors.

\begin{figure}[t]
  \includegraphics[height=.27\textheight]{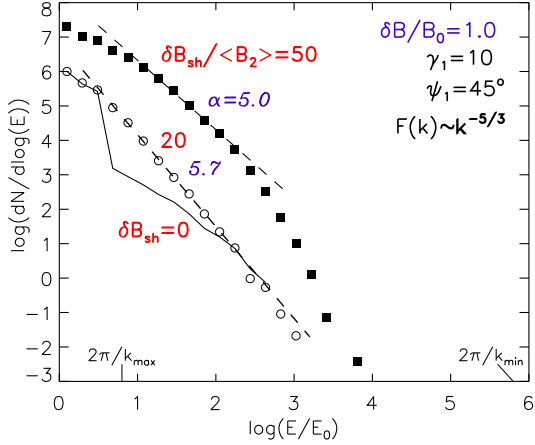}
  \caption{Particle spectra for superluminal 
  shocks with $\gamma_1$=10 formed in 
the presence of shock-generated downstream turbulence and the Kolmogorov
power spectrum of the background field component (solid line --- compare Fig.~1).
The amplitudes of the short-wave perturbations, $\delta B_{sh}/\langle B_2\rangle$, are 
given near the respective spectra.}
\end{figure}

The effects of the turbulent field compression may also occur in parallel 
high-$\gamma$ shocks (Fig. 2) for large-amplitude perturbations. In this case, the 
field compression leads to an effectively perpendicular shock configuration,
and features analogous to those observed in oblique shocks are 
recovered. Only for weakly perturbed magnetic fields can the 
wide-energy range particle spectra be formed. However, they are non-power-law in 
the full energy range, and their power-law parts are flat ($\alpha < 4$) due to 
the effects of long-wave perturbations. The convergence of the spectra to the
``universal'' spectral index ($\alpha\approx 4.2$) claimed in the literature 
[e.g., 1, 3, 4] %\cite{bed98,ach01,kir00} 
is clearly not observed.

\begin{figure}[t]
\begin{center}
  \includegraphics[height=.25\textheight]{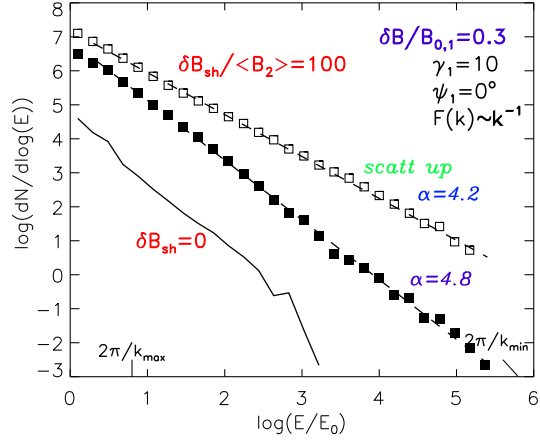}
\end{center}
  \caption{Accelerated particle spectra at parallel shocks with $\gamma_1$=10 in 
the presence of shock-generated turbulence and weakly perturbed background field. 
The spectrum indicated as 
``{\it scatt up}" was obtained in the {\em unphysical} model in, which the particle 
pitch-angle diffusion was
assumed to exist both downstream and upstream of the shock.}
\end{figure}
More realistic microscopic models of collisionless shocks show that the shocks
can generate a highly nonlinear {\em short-wave} turbulence downstream due to 
filamentation instabilities at the shock front \cite[e.g.]{fre04}. 
In \cite{nie06b}
we augmented the magnetic field model by this shock-generated
component, which can provide efficient particle scattering and may lead to a 
decorrelation between particle motion and the compressed field downstream of the 
shock. 
For the case of oblique shocks (Fig. 3), increasing the amplitude of the 
shock-generated turbulence 
leads to a more efficient acceleration with particle spectral tails extending 
to higher energies. However, in all cases, in which  
$\delta B_{sh}/\langle B_2\rangle \gg 1$, the energetic spectral tails are convex, 
and the spectra have cutoffs at energies for which the resonance
condition for interactions with compressed turbulence is fulfilled. 
This is because the influence on particle trajectories
of the shock-generated turbulence decreases with increasing particle energy,
and eventually becomes smaller than the influence of the large-scale
background field. Similar spectral effects are also observed for parallel shocks
when the amplitude of the long-wave background component is large. 
Extended power-law particle distributions can be formed in parallel shocks 
propagating in a medium with low-amplitude of the long-wave turbulence (Fig. 4). 
However, the spectra are steeper than the expected ``universal'' spectrum,
$\alpha > \alpha_u$. The only case in
which we were able to obtain spectra with $\alpha = \alpha_u$ in the energy
range considered, involved the
{\em unphysical} model with the short-wave component introduced both downstream and 
{\it upstream} of the shock (spectrum indicated as ``{\it scatt up}" in Fig. 4),
which removed the effects of upstream long-wave perturbations.
\subsection{Discussion and conclusions}
Our results require a revision of many 
earlier discussions of cosmic-ray acceleration up to very high energies
in the first-order Fermi process at relativistic shocks. 
The modeling shows that turbulence  
consistent with the shock jump conditions can lead to a
substantial modifications of the acceleration picture as compared to  
simplified models producing wide-range power-law energy distributions, often 
with the ``universal'' spectral index \cite{bed98,ach01,kir00}. 
The presence of highly nonlinear short-wave
turbulence at the shock can lead to more efficient acceleration, but the amplitude
of the shock-generated component required to produce extended power-law spectra 
is unrealistically high, in particular for large shock Lorentz factors.
Our simulations show 
that relativistic shocks, being essentially always superluminal, 
possibly generate accelerated particle distributions with cutoffs below either 
the maximum resonance energy enabled by the {\it high-amplitude} background 
turbulence ($r_g(E_{cutoff}) < \lambda(E_{res,max})$), or approximately at the 
energy of the compressed background plasma ions 
$E_{cutoff} \sim \gamma_1 m_{i}c^2$. Thus, in conclusion, relativistic shocks 
are not promising sites as possible sources of  
ultra--high-energy cosmic rays. Should UHECR production be expected from
relativistic shocks it must invoke other processes, e.g. the second-order Fermi
process in downstream relativistic MHD turbulence \cite{vir05}. 

Finally let us note that our models might have recently acquired observational
confirmation. 
The recent {\em Spitzer} imaging of Cygnus A hotspots 
resulted in the detection of the high-energy tails of their  
synchrotron  radiation \cite{staw07}.  
Combined with data collected at other frequencies,
these observations allowed for a detailed modeling of the broad-band emission 
from the two brightest 
hotspots, which put precise constraints on the underlying 
energy spectra of ultrarelativistic electrons. The spectra can be approximated
by a broken power-law with the flat low-energy spectral index $\alpha_l\approx 3.5$
followed by a steep high-energy part with $\alpha_h > 5$, with the break energy 
corresponding approximately to the proton rest mass energy. Thus, the shape of the spectra reflects
most likely two different regimes of the electron acceleration process at 
mildly relativistic shocks of the hotspots: the preacceleration processes responsible
for the spectral shape below the critical energy scale given by the inertia of
protons, above which the first-order Fermi process operates. The steep slope 
of the spectra at high-energies is therefore in agreement with our 
modeling of the Fermi processes at oblique mildly relativistic shocks. In fact,
the differences in the high-energy power-law indices and cut-off energies observed 
between the two hotspots may be attributed to the sensitivity of the Fermi process
to the measured differences in the intensity (and possibly configuration) of the 
magnetic field at the shocks in the hotspots.

{\small \noindent 
This work was supported by MNiSW in years 2005-2008 as a research project 
1 P03D 003 29.}
%
%This is the reference to .bib file (Whitout .bib!)
\bibliography{icrc1051}
%This in the bibtex style, is ok.
\bibliographystyle{plain}

\end{document}